\documentclass[12pt]{article}
\usepackage[a4paper,margin=2cm]{geometry}
\usepackage{amsmath,amsthm} 
\usepackage[english]{babel} 
\usepackage{epsfig} 
\usepackage{parskip}
\usepackage{setspace}
\usepackage{eurosym}
\newcommand{\bc}{\begin{center}}
\newcommand{\ec}{\end{center}}

\begin{document}
\pagestyle{plain}
\singlespacing
\begin{flushright}
\small{Proceedings of the 7$^{th}$ International Symposium on Radiative Transfer, RAD-13\\
June 2-8, 2013, Ku\c{s}adasi, Turkey\\
\ \\
RAD-13-240}\\
\end{flushright}
\ \\

\thispagestyle{empty}
\begin{center}
\textbf{Energy density above a resonant metamaterial in the GHz : an alternative to near-field thermal emission detection.}\\
\ \\
Karl Joulain, J\'er\'emie Drevillon and Youn\`es Ezzahri$^{*}$\\
$^{*}$Institut Pprime, Universit\'e de Poitiers-CNRS-ISAE-ENSMA, BP 633\\
86022 Poitiers Cedex, France
\end{center}
\ \\

ABSTRACT. This paper proposes an experiment to easily detect radiative heat transfer in the microwave range. Following an idea given by Pendry more than a decade ago \cite{Pendry:1998}, we show that a 3D array of tungsten 2$\mu$m radius wires with a 1 cm period makes a low cost material exhibiting a surface plasmon in the microwave range around 2.9 GHz. Such a heated material should exhibit an emission peak  near the plasmon frequency well above ambient emission. Analysis of  the signal detected in the near-field should also be a tool to analyze how homogenization theory applies when the distance to the material is of the order of the metamaterial period. It could also be give a model to non-local dielectric properties in the same conditions.

\bc\textbf{INTRODUCTION}\ec 

Since the seminal works of Rytov \cite{Rytov:1989ur}, Polder and Van Hove \cite{Polder:1971uu}, Agarwal \cite{Agarwal:1975ur}, it is known that energy density can be enhanced from vacuum energy density when the distance involved is lower than the wavelength. This energy density is actually a measure of the density of electromagnetic states (LDOS) \cite{Joulain:2003hc} so that a large increase of the LDOS close to a surface implies an increase of the thermal energy density. Such an increase of the density of energy occur when the material support surface waves such as plasmon polariton or phonon polariton \cite{Shchegrov:2000td,Joulain:2005tz}. This enhancement in the energy density has been recently detected by two types of methods. Above a dielectric like SiC, near-field infrared radiation has been detected using near-field radiation scattering by a metallic probe leading to a new technique named Thermal Radiation Scanning Tunneling Microscopy (TRSTM) \cite{DeWilde:2006kt}. Other experiments  measuring a heat transfer between a sample and a probe above a metal \cite{Kittel:2005fr} or a dielectric \cite{Rousseau:2009es,Narayanaswamy:2008gj} have shown near-field thermal radiation enhancement. However, such measurements can hardly be done on the spectral domain although some of them seems to have been performed recently \cite{Raschke,Dewilde}.

An alternative to such measurements that necessitate to act at the submicronic scale, is to make measurements in the microwave domain. In principle, there are two advantages to work at such low frequencies. First of all, if we have for instance a resonance in the Ghz i.e for a typical wavelength of 10 cm, this resonance should be seen at distances to the material of the order of 1 cm. No needs of specific positioning system is therefore required. Moreover, detection system in the GHz are able to detect and follow the electromagnetic field. This means that an emission spectrum can easily be achieved. The problem is that resonant surface waves usually occur in the visible domain for metals supporting surface plasmon and in the infrared domain for dielectric supporting surface polariton. We suggest here to use a 3D centimetric cubic array of micrometric tungsten wires. Such an array should produce a surface plasmon around a frequency of 3 GHz and exhibit a resonant peak in the density of near-field thermal electromagnetic energy around this frequency. Note that our work follows the pioneering ideas developped at Institut Fresnel in Marseille to study radiative properties of complex shape particles. Instead of directly study in the visible or the infrared range scattering by complex particles or aggregates, these particles are studied in the microwave range where they size have been increased in order to keep the ratio size over wavelength constant \cite{Merchiers:2009,Vaillon2011,Geffrin2012}.

In the following, we remind how to calculate effective optical properties of a cubic array made of metallic wires. We propose to build such a structure and then use these optical properties in order to calculate energy density above such a structure. We show that a significative enhancement should be detected above these structures at centimetric distances.

\bc\textbf{METAMATERIAL SAMPLE}\ec 
\begin{figure}[tb]
\centering
\includegraphics[width=5in]{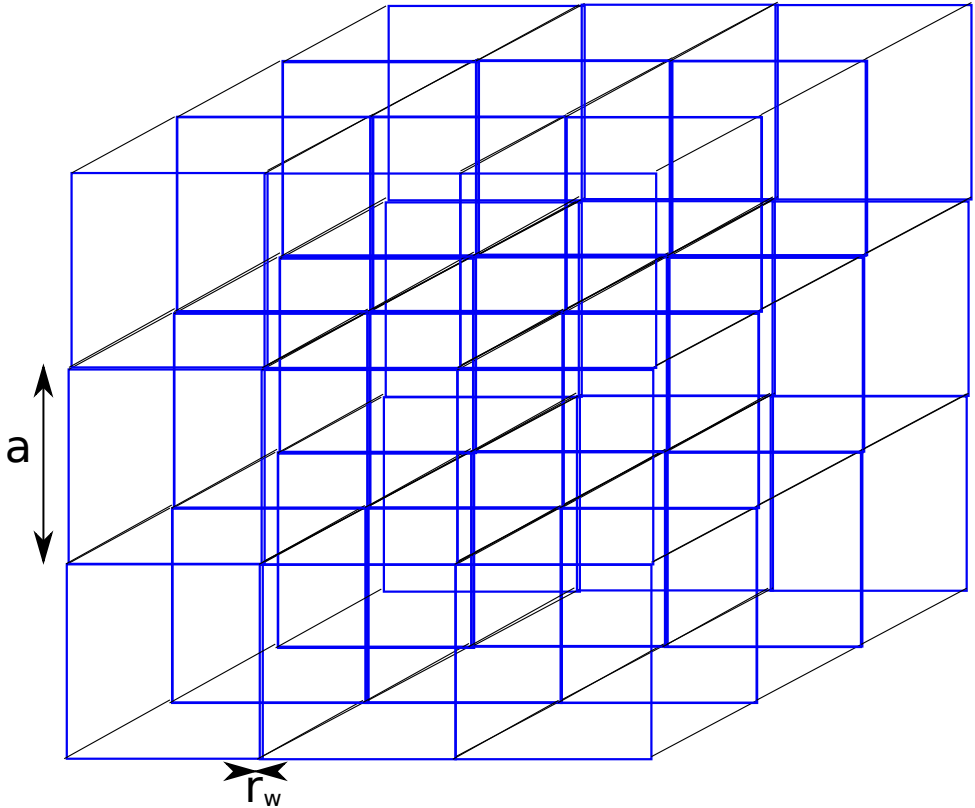}
\caption{Metamaterial made of a 3-dimensional cubic array of cylindrical wires. The wires (here tungsten wires) are separated by $a$ in the three directions and have all the same radius $r_w$.}
\label{fig:sample}
\end{figure}

\begin{figure}[tb]
\centering
\includegraphics[width=3in]{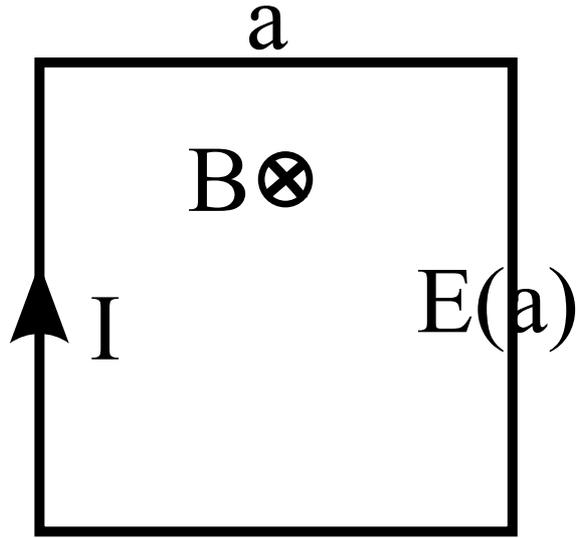}
\caption{Elementary frame of metamaterial. The frame is a square which side is $a$. The current $I$ circulates in the frame. The magnetic field $B$ created by the current is in the perpendicular direction to the frame. The electric field $E(a)$ is the electric field in the right part of the frame at a distance $a$ from the left part of the frame.}
\label{fig:frame}
\end{figure}

Let us consider a structure a cubic array made of metallic wires (Fig.\ref{fig:sample}). We reproduce here a calculation originally made by Pendry \cite{Pendry:1998} but giving more details than in the original article. Let us denote $r_w$ the wire radius and $a$ the cubic array period. An elementary wire constituting the array and crossed by a current $I$ produces a magnetic field $H$ in the orthoradial direction of the wire. The magnetic field, given in any textbook reads
\begin{equation}
\label{ }
H=\frac{B}{\mu_0}=\frac{I}{2\pi r}
\end{equation}
Here, $B$ is the magnetic induction, whereas $\mu_0$ is the vacuum permeability and $r$ the distance to the wire axis. Let us now apply the Maxwell-Faraday equation to an elementary frame constituting the array (Fig. \ref{fig:frame}). The circulation of the electric field is equal to the time derivative of the flux of the magnetic induction through the elementary frame. Note here that the time derivative of a quantity will be obtained by a simple multiplication by $-i\omega$ since the quantity considered will be monochromatic with a time dependence chosen as $e^{-i\omega t}$. Let us first calculate magnetic induction flux $\Phi$ through an elementary frame
\begin{equation}
\label{ }
\Phi=\frac{\mu_0Ia}{2\pi}\int_{r_w}^{a}\frac{dr}{r}=\frac{\mu_0Ia}{2\pi}\ln\frac{a}{r_w}
\end{equation}

So that the induction $L$ of an elementary frame can be written $L=\mu_0a/(2\pi)\ln a/r_w$. Maxwell-Faraday equation around the frame reads
\begin{equation}
\label{IdeE}
RI-aE(a)=-\frac{d\Phi}{dt}=-L\frac{dI}{dt}=i\omega LI
\end{equation} 
where $R$ is the resistance of the wire of length $a$. In terms of the conductivity, this resistance expresses $R=a/(2\pi\sigma r_w^2)$.
We now suppose that the wavelength involved is much smaller than the array period $a$ so that $\lambda\ll a$, then the electric field in the structure will not vary a lot at the scale of the elementary frame. The preceding equation therefore gives us a first relation between the current and the electric field opening the way to the determination of an effective dielectric constant for the material. Indeed, dielectric constant in a medium relates polarization to the electric field and the volumetric current is related to the polarization through $\mathbf{j}=\partial\mathbf{P}/\partial t=-i\omega\mathbf{P}=-i\omega\epsilon_0(\epsilon_r-1)\mathbf{E}$.  Integrating this relation on an elementary cube constituting the array
\begin{equation}
\label{IdeEeps}
aI=-i\omega\epsilon_0(\epsilon_r-1)\mathbf{E}a^3
\end{equation}
Combining equations (\ref{IdeE}) and (\ref{IdeEeps}), one obtains the metamaterial effective dielectric constant
\begin{equation}
\label{ }
\epsilon_r=1-\frac{\frac{1}{\epsilon_0aL}}{\omega^2+i\omega R/L}=1-\frac{\Omega_p^2}{\omega(\omega+i\gamma)}
\end{equation}
The right expression is a Drude expression where the plasma frequency can be identified with
\begin{equation}
\label{ }
\Omega_p^2=\frac{2\pi c^2}{a^2\ln(a/r_w)}
\end{equation}
and the damping frequency with
\begin{equation}
\label{ }
\gamma=\frac{R}{L}=\frac{\epsilon_0 a^2 \Omega_p^2}{\pi r_w^2\sigma}
\end{equation}
Plasma frequency expression show that it is of the order of $c/a$ with a very slow correction depending on the ration $a/r_w$. The larger this ratio will be, the smaller will be the plasma frequency and the larger will be the plasma wavelength $\lambda_p=2\pi c/\Omega_p$. Resonant phenomena occur near the plasma frequency and the plasma wavelength. It is important to have this wavelength as large as possible in order that it is larger than the array period. Indeed, near-field effects occurs at distances smaller than the wavelength. However, this theory is valid only if the medium can be seen as an effective medium which can hardly be the case if the distance to the material is of the order of the metamaterial period.

Such a metamaterial could be achieved with tungsten (W) wires. Very small wires ($r_w=2\times10^{-6}$ m) can be found and sold by companies at a price of around 750 \euro $ $ for a 100 m-long coil. Let us consider an array of period $a = 1$ cm, then $\Omega_p= 2.58\times10^{10}$ rad s$^{-1}$ so that $\lambda_p=7.31$ cm. The losses in this model is obtained from the a Drude model fitted on the data measured by Ordal et al. \cite{Ordal}.
\begin{equation}
\label{ }
\epsilon_W=1-\frac{\omega_p^2}{\omega(\omega+ i \omega_\tau)}
\end{equation}
where $\omega_p=9.12\times 10^{15}$ rad s$^{-1}$ and $\omega_\tau=8.16\times10^{13}$ rad s$^{-1}$. Knowing that at low frequency, the dielectric constant is related to the electrical conductivity $\sigma$ by $\epsilon_W=1+i\sigma/(\omega\epsilon_0)$, the electrical conductivity is then given by
$\sigma=\epsilon_0\omega_p^2/\omega_\tau=5.67\times 10^7$ $\Omega^{-1}$ m$^{-1}$. In terms of $\gamma$,  the losses in the metamaterial are $\gamma = $ 8.26$\times$10$^8$ rad s$^{-1}$. As it has been seen in the past and as it will be seen in the next section, the LDOS and energy density enhancement is the most important near the frequency at which the material support a surface polariton. This happens when the dielectric constant approaches -1, that is $\omega=\Omega_p/\sqrt{2}$.  At this frequency, the field penetration depth $\delta$ will be such that $\Im(n)\omega\delta/c=1$ where $n=\sqrt(\epsilon)$ is the optical index. If $\epsilon=-1$, $n=i$ so that $\delta=a\sqrt{2\ln(a/r_w))/\pi}=  2.32$ cm. Note that the wavelength where the resonance occurs is $\lambda=\sqrt{2}\lambda_p= 10.34$ cm that is a frequency $\nu=2.90$ GHz.

If we want a structure which lateral size is larger than the wavelength and which thickness is of a few $\delta$ in order to be considered as an opaque body, we need, as a minimum structure a parallelepiped of 20 cm $\times$ 20 cm $\times$ 5 cm. This structure will contain 2000 elementary cubes each of them being constituted of 3 cm of wire (note that each wire is shared between 4 cubes and that a cube is constituted by 12 1cm-long wires). This structure will therefore need 60 m of wires and can be achieved in a 100 m-long tungsten coil cited above.

\bc\textbf{NEAR-FIELD THERMAL ENERGY DENSITY ABOVE THE METAMATERIAL}\ec 

We now calculate the energy density above a semi-infinite metamaterial made of micrometric tungsten wires described in the preceding section. This metamaterial is now described as an homogeneous model which dielectric properties obey a Drude model. Metamaterial temperature is $T$ and the distance of the metamaterial at which the energy density is calculated is $z$. We used the fluctuationnal electrodynamics theory developped by Rytov \cite{Rytov:1989ur}. Energy density is then given by an integration over plane waves ($\mathbf{K},k_{zv}$) where $\mathbf{K}$ is the parallel wave vector and $k_{zv}$ the wave vector perpendicular to the interface in the vacuum such that $K^2+k_{zv}=k_0^2=\omega^2/c^2$.
\begin{eqnarray}
u(\omega,z,T) & = & \frac{\Theta(\omega,T)\omega^2}{2\pi^2c^3}\left\{\int_0^{\omega/c}\frac{KdK}{k_0|k_{zv}|}\frac{(1-|r^s|^2)+(1-|r^p|^2)}{2} \right.\nonumber\\
 & + & \left.\int_{\omega/c}^\infty \frac{4K^3dK}{k_0^3|k_{zv}|}\frac{\Im(r^s)+\Im(r^p)}{2}e^{-2\Im(k_{zv})z} \right\} \label{udens}
\end{eqnarray}
where $\Theta(\omega,T)=\hbar\omega/(e^{\hbar\omega/k_bT}-1)$ is the mean energy of an oscillator at frequency $\omega$ and temperature $T$. $r^s$ and $r^p$ are the Fresnel coefficients given by
\begin{eqnarray}
r^s & = & \frac{k_{zv}-k_z}{k_{zv}+k_z} \\
r^p & = & \frac{\epsilon_rk_{zv}-k_z}{\epsilon_rk_{zv}+k_z} 
\end{eqnarray}
where $k_z=\sqrt{\epsilon_rk_0^2-K^2}$.
In expression (\ref{udens}), the first integral is over wave vector which parallel wave vector is lower than $k_0$ and corresponds to propagative waves. The second integral over parallel wave vector larger than $k_0$ describes the contribution of evanescent waves which are exponentially decaying from the interface. This last contribution can only be seen in the near field at sub wavelength distances. It is important when the imaginary part of the Fresnel reflection coefficient is large. This happens in $p$-polarization near the polariton surface wave frequency where $\epsilon_W=-1$ \cite{Joulain:2005tz} 

\begin{figure}[tb]
\centering
\includegraphics[width=4in]{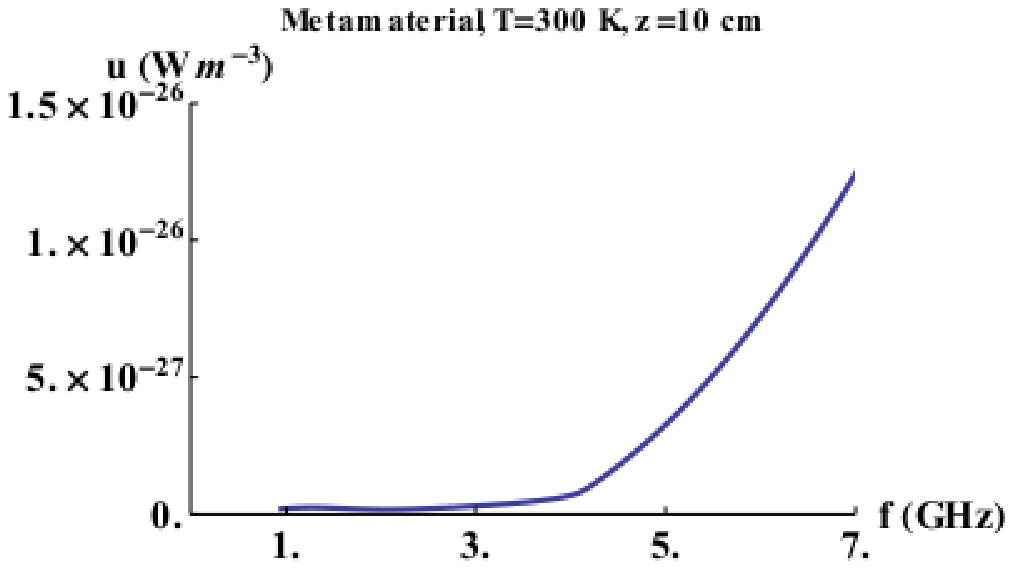}
\includegraphics[width=4in]{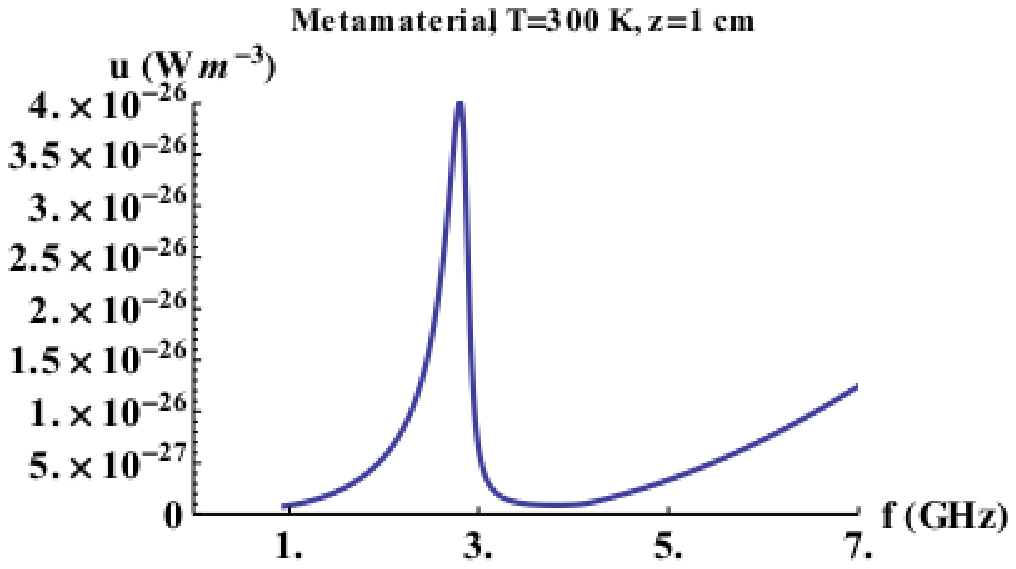}
\includegraphics[width=4in]{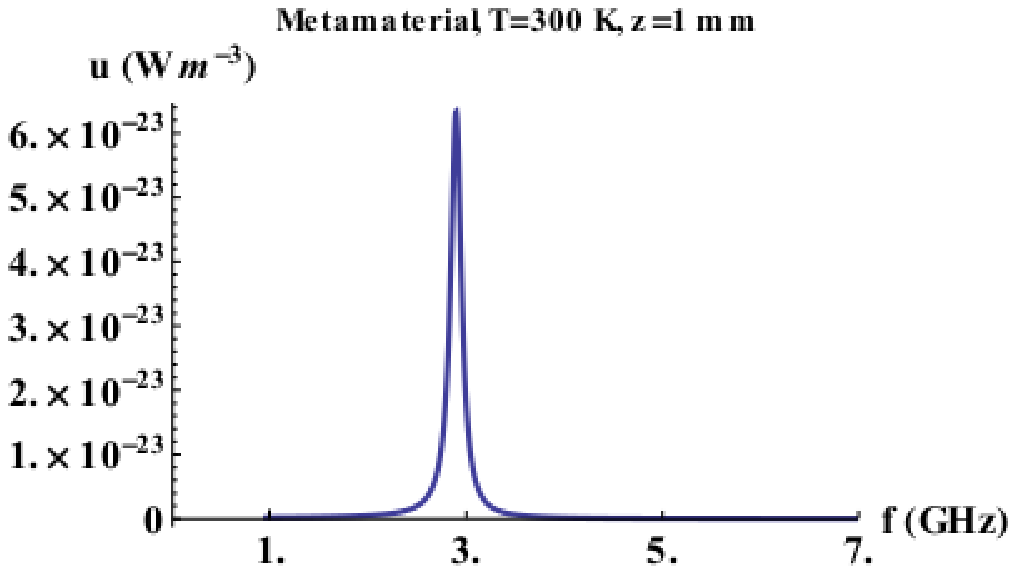}
\caption{Density of energy versus frequency above a metamaterial made of 2 micrometer radius tungsten wires with a 1 cm period. The density of energy is calculated for three distances, 10 cm (top figure), 1 cm (middle figure) and 1 mm (bottom figure). At small distance a peak appears in the density of energy at the frequency where the metamaterial exhibit a surface polarities (here 2.9 GHz).}
\label{fig:u}
\end{figure}

In Fig. \ref{fig:u}, we have represented the density of energy above the metamaterial for 3 different distances. We first calculated this quantity at a distance of 10 cm from the semi-infinite metamaterial. Actually, at this distance, the energy density is the same as the one we would find in far-field.  The density of energy increases with the frequency around 3 GHz because, this part of the spectrum is situated in the Rayleigh-Jeans approximation of the Planck's law where the density of energy increases as the square of the frequency.

At 1 cm from the interface, the density of energy is completely different from the preceding case. $u$ exhibits a peak around 3 GHz. We see that around the peak, the density of energy has increased by more than one order of magnitude. The peak is situated at 2.9 GHz i.e. for $\omega=1.82\times10^{10}$ rad s$^{-1}$. It has a width of the order of the metamaterial damping parameter $\gamma$. The resonant frequency correspond to the frequency where the reflection coefficient in $p$-polarization diverges that is when the dielectric constant approaches -1. At this frequency, the metamaterial exhibit a surface polariton which is here a surface plasmon. In the same way it has been seen for materials supporting surface polariton such as polar material like SiC or SiO2, the metamaterial exhibits a peak in the density of energy that is also a peak in the density of electromagnetic states. 

If the density of energy is measured at a distance of 1 mm, we note that the peak is greatly enhanced and that the density of energy has increased by three orders of magnitude. At this distance the theory used here is probably note valid because the distance to the material is much smaller than the distance between the wires constituting the metamaterial. At this scale, one can hardly see the metamaterial as a continuous medium. Indeed, at 2.9 GHz, the wavelength involved is of the order of 10 cm which can be considered as large compare to the material period. This means that when the plane waves involved in the calculation are mainly propagative waves (i.e waves for which $K<\omega/c$), spatial variation along the interface are of the order of  $2\pi/K>\lambda$ and larger than the wavelength.  When one approaches the metamaterial, evanescent waves begin to enter into play. Typically, the largest parallel wave vector involved are of the order of $2\pi/z$. This means that spatial variation along the interface will also be of the order of $z$. Therefore, $z\approx a$ is the smallest distance for which one can consider that all the contributing evanescent waves have a lateral spatial variation smaller than the metamaterial period. Therefore, one can consider our calculation as a plausible approximation as long as $z\geq a$. When $z$ is smaller than $a$, it is hard to say at which distance the effective medium theory will lose its pertinence. Note however that as long as the theory applies, i.e. for distance as small as 1 cm, the effective medium calculation is robust and is not very sensible to the metamaterial irregularities. The reasoning leading to the metamaterial optical response presented above can be written in any elementary frame even if the material is not strictly periodic. What is important is the average wire radius $r_w$ as well as the average frame size $a$. A variation of a few percent in the periodicity $a$ should affect the optical response parameters in the same orders since the variations with $a$ and $r_w$ follow a power or logarithmic law.

We think an experiment detecting the energy density at 3 GHz could help to determine in which conditions effective medium theory could apply. This experiment would consist in a small antenna or detector that would be approached to the metamaterial and that would measure the square modulus of the electric field that is a quantity proportional to the electromagnetic density of energy. We also think this experiment could help to understand what are the limits of fluctuationnal electrodynamics with a local approximation for the materials modeling at very short distance. Indeed, for local media, i.e for media which do not exhibit spatial dispersion, the energy energy density diverges as $1/z^3$ when one approaches a heated body. Moreover, when two semi-infinite heated bodies are put at small distance one from each other, it can be seen that the $z$-dependence in the heat transfer between the bodies is $1/z^2$. This divergence is of course unphysical for extremely short distance where non-local effects come into play. Such local effects have been studied in the case of metals \cite{Chapuis:2008kca}. The Thomas-Fermi length acts as a cut-off length preventing the divergence at subnanometric distances.  For dielectrics, it has been suggested that the lattice spacing plays a similar role \cite{Henkel:2006iqa}. The experiment measuring the increasing of the electromagnetic energy density should show at which distance the increasing stops and where the saturation begins. This should give informations about the scale below which local approximation cease to be valid. One could then extrapolate this behavior to the case of usual matter and see if it is relevant to set the locality length to lattice constant.

\bc\textbf{CONCLUSION}\ec 

We have shown in this article that a metamaterial consisting of a regular centimetric cubic array of 2 micrometer-radius tungsten wires exhibits surface polaritons in the microwave range around 2.9 GHz. Such a metamaterial, that can be build at a low cost price, will exhibit when heated a peak in the near-field thermally emitted density of enegy at a typical 1 cm distance from the material. These near-field experiment are in principle much more easier to realize than in the visible rage because they do not need sophisticated nanopositionning system. Making a study of the density of energy variation with the distance should help to understand two questions that remain without answer at present time : at what scale the homogeneous approximation to describe a metamaterial remains valid? At which distance from a material local approximations describing bulk materials are still valid and what is the local length that enter into play at extreme distances from a material.


\begin{thebibliography}{10}

\bibitem{Pendry:1998}
JB~Pendry, AJ~Holden, DJ~Robbins, and WJ~Stewart,
\newblock ``{Low frequency plasmons in thin-wire structures}'',
\newblock {\em {Journal of Physics-Condensed Matter}}, vol. {10}, no. {22}, pp.
  {4785--4809}, {1998}.

\bibitem{Rytov:1989ur}
S.M. Rytov, Y.A. Kravtsov, and V.T. Tatarskii,
\newblock {\em {Principle of Statistical Radiophysics 3}}, vol.~3 of {\em
  Elements of Radiation Fields},
\newblock Springer Verlag, 1989.

\bibitem{Polder:1971uu}
D~Polder and M.~van Hove,
\newblock ``{Theory of Radiative Heat Transfer between Closely Spaced Bodies
  }'',
\newblock {\em Physical Review B}, vol. 4, pp. 3303--3314, Nov. 1971.

\bibitem{Agarwal:1975ur}
G.S. Agarwal,
\newblock ``{Quantum electrodynamics in the presence of dielectrics and
  conductors. I. Electromagnetic-field response functions and black-body
  fluctuations in finite geometries}'',
\newblock {\em Physical Review A}, vol. 11, pp. 230--242, Jan. 1975.

\bibitem{Joulain:2003hc}
Karl Joulain, R{\'e}mi Carminati, Jean-Philippe Mulet, and Jean-Jacques
  Greffet,
\newblock ``{Definition and measurement of the local density of electromagnetic
  states close to an interface}'',
\newblock {\em Physical Review B}, vol. 68, no. 24, pp. 245405, Dec. 2003.

\bibitem{Shchegrov:2000td}
A.V. Shchegrov, K~Joulain, R~Carminati, and J~J Greffet,
\newblock ``{Near-field Spectral Effects due to Electromagnetic Surface
  Excitations}'',
\newblock {\em Physical Review Letters}, vol. 85, pp. 1548--1551, Aug. 2000.

\bibitem{Joulain:2005tz}
Karl Joulain, Jean-Philippe Mulet, , , and Jean-Jacques Greffet,
\newblock ``{Surface Electromagnetic Waves Thermally Excited: Radiative Heat
  Transfer, Coherence Properties and Casimir Forces Revisited in the Near
  Field}'',
\newblock {\em arXiv.org}, vol. physics.optics, Apr. 2005.

\bibitem{DeWilde:2006kt}
Yannick De~Wilde, Florian Formanek, R{\'e}mi Carminati, Boris Gralak,
  Paul-Arthur Lemoine, Karl Joulain, Jean-Philippe Mulet, Yong Chen, and
  Jean-Jacques Greffet,
\newblock ``{Thermal radiation scanning tunnelling microscopy}'',
\newblock {\em Nature}, vol. 444, no. 7120, pp. 740--743, Dec. 2006.

\bibitem{Kittel:2005fr}
Achim Kittel, Wolfgang M{\"u}ller-Hirsch, J{\"u}rgen Parisi, Svend-Age Biehs,
  Daniel Reddig, and Martin Holthaus,
\newblock ``{Near-Field Heat Transfer in a Scanning Thermal Microscope}'',
\newblock {\em Physical Review Letters}, vol. 95, no. 22, pp. 224301, Nov.
  2005.

\bibitem{Rousseau:2009es}
Emmanuel Rousseau, Alessandro Siria, Guillaume Jourdan, Sebastian Volz, Fabio
  Comin, Jo{\"e}l Chevrier, and Jean-Jacques Greffet,
\newblock ``{Radiative heat transfer at the nanoscale}'',
\newblock {\em Nature Photonics}, vol. 3, pp. 514--517, Aug. 2009.

\bibitem{Narayanaswamy:2008gj}
Arvind Narayanaswamy, Sheng Shen, and Gang Chen,
\newblock ``{Near-field radiative heat transfer between a sphere and a
  substrate}'',
\newblock {\em Physical Review B}, vol. 78, no. 11, pp. 115303, Sept. 2008.

\bibitem{Raschke}
Andrew~C. Jones and Markus~B. Raschke,
\newblock ``{Thermal Infrared Near-Field Spectroscopy}'',
\newblock {\em {Nano Letters}}, vol. {12}, no. {3}, pp. {1475--1481}, {2012}.

\bibitem{Dewilde}
Arthur Babuty, Karl Joulain, Pierre-Olivier Chapuis, Jean-Jacques Greffet, and
  De~Wilde Yannick,
\newblock ``{Blackbody spectrum revisited in the Near Field}'',
\newblock {\em {Phys. Rev. Lett}}, vol. {110}, pp. {146103}, {2013}.

\bibitem{Merchiers:2009}
Olivier Merchiers, Jean-Michel Geffrin, Rodolphe Vaillon, Pierre Sabouroux, and
  Bernard Lacroix,
\newblock ``{Microwave analog to light scattering measurements on a fully
  characterized complex aggregate}'',
\newblock {\em {Applied Physics Letters}}, vol. {94}, no. {18}, {2009}.

\bibitem{Vaillon2011}
Rodolphe Vaillon, Jean-Michel Geffrin, Christelle Eyraud, Olivier Merchiers, Pierre Sabouroux and Bernard Lacroix
\newblock ``{A new implementation of a microwave analog to light scattering measurement device}'',
\newblock {\em {Journal of Quantitative Spectroscopy and Radiative Transfer}}, vol. {112}, pp. {1753-1760}, {2011}.

\bibitem{Geffrin2012}
J.-M. Geffrin, B. Garcia-Camara, R. Gomez-Medina, P. Albela, L.S. Froufe-Perez, C. Eyraud, A. Litman, R. Vaillon, F. Gonzalez, M. Nieto-Vesperinas, J.J. Saenz and  F. Moreno.
\newblock ``{Magnetic and electric coherence in forward- and back-scattered electromagnetic waves by a single dielectric subwavelength sphere}'',
\newblock {\em {Nature Communications}}, vol. {3}, pp. {1171-1178}, {2012}.



\bibitem{Ordal}
MA~Ordal, RJ~Bell, RW~Alexander, LA~Newquist, and MR~Querry,
\newblock ``{Optical-Properties of Al, Fe, Ti, Ta, W, and Mo at Submillimeter
  Wavelengths}'',
\newblock {\em {Applied Optics}}, vol. {27}, no. {6}, pp. {1203--1208}, {1988}.

\bibitem{Chapuis:2008kca}
Pierre-Olivier Chapuis, Sebastian Volz, Carsten Henkel, Karl Joulain, and
  Jean-Jacques Greffet,
\newblock ``{Effects of spatial dispersion in near-field radiative heat
  transfer between two parallel metallic surfaces}'',
\newblock {\em Physical Review B}, vol. 77, no. 3, pp. 035431, Jan. 2008.

\bibitem{Henkel:2006iqa}
C~Henkel and K~Joulain,
\newblock ``{Electromagnetic field correlations near a surface with a nonlocal
  optical response}'',
\newblock {\em Applied Physics B}, vol. 84, no. 1-2, pp. 61--68, Apr. 2006.

\end{thebibliography}
\end{document}